\documentclass[12pt]{iopart}   % style for IOP journals

\usepackage{graphicx}   % for figures
\usepackage{graphics}% Include figure files
\usepackage{dcolumn}% Align table columns on decimals point
\usepackage{bm}% bold math

\usepackage{multirow}

\usepackage{latexsym}
\usepackage{amssymb}

\usepackage{color}
\usepackage{epstopdf}

\begin{document}

\title[Microscopic description of twisted magnetic Cu$_2$OSeO$_3$]{Microscopic description of twisted magnetic Cu$_2$OSeO$_3$}
\author{Viacheslav A Chizhikov and Vladimir E Dmitrienko}
\address{A V Shubnikov Institute of Crystallography, Leninski\u{\i} pr. 59, 119333 Moscow, Russia}
\ead{chizhikov@crys.ras.ru, dmitrien@crys.ras.ru}

\begin{abstract}
Twisted structures of chiral cubic ferromagnetics MnSi and Cu$_2$OSeO$_3$ can be described both in the frame of the phenomenological Ginzburg--Landau theory and using the microscopical Heisenberg formalism with a chirality brought in by the Dzyaloshinskii--Moriya (DM) interaction. Recent progress in quantum first-principal methods allows to calculate interatomic bond parameters of the Heisenberg model, namely, isotropic exchange constants $J_{ij}$ and DM vectors $\mathbf{D}_{ij}$, which can be used for simulations of observed magnetic textures and comparison of their calculated characteristics, such as magnetic helix sense and pitch, with the experimental data. In the present work, it is found that unaveraged microscopical details of the spin structures (the local canting) have a strong impact on the global twist and can notably change the helix propagation number. Coefficients ${\cal J}$ and ${\cal D}$ of the phenomenological theory and helix propagation number $k={\cal D}/2{\cal J}$ are derived from interatomic parameters $J_{ij}$ and $\mathbf{D}_{ij}$ of individual bonds for MnSi and Cu$_2$OSeO$_3$ crystals and similar cubic magnetics with almost collinear spins.
\end{abstract}

\pacs{75.25.-j, 75.50.Gg, 75.85.+t, 75.10.Hk}
\submitto{\JPCM}
\maketitle
%\Large

\section{Introduction}
\label{sec:intro}

The magnetic properties of
the cubic crystal Cu$_2$OSeO$_3$ are of great interest for several
reasons. First of all, having the space group $P2_13$ without centre of
inversion, this crystal becomes a cubic helimagnet below the critical
temperature of about 58 K. Moreover, it is the first cubic crystal beyond
the class of itinerant magnetics with $B20$ crystal structure
\cite{Grigoriev2006a,Muhlbauer2009,Munzer2010,Adams2011}, for which
A-phase, associated with Skyrmion lattice \cite{Bogdanov89,Rossler06,Ambrose2013}, has
been recently observed
\cite{SekiSci2012,AdamsPRL2012,SekiPRB2012,OnosePRL2012,SekiPRB2012-2,White2012}.
Secondly, being an insulator, Cu$_2$OSeO$_3$ has magnetoelectric
properties
\cite{White2012,Bos2008,Miller2010,Maisuradze2011,Maisuradze2012,Belesi2012,GongXiang2012}, which provides a new physical significance in comparison with the $B20$ crystals.
The interconnection between magnetization gradients and electric
polarization makes this crystal potentially applicable for data storing
devices and spintronics \cite{Liu2013,Liu2013-2,Jennings2013,PyatakovUFN2012}. And thirdly, but
not finally, the more complex than $B20$ structure (16 magnetic copper
atoms in two non-equivalent positions in the unit cell) makes it an
interesting object for studying the spin textures.

Most of the known twisted magnetics possess symmetry lower than cubic.
Their strong anisotropy orients magnetic helices in special
crystallographic directions. As opposed to them, in cubic magnetics
without centre of inversion, the binding between magnetic and crystal
structures can be so subtle, that the helix can be easily reoriented along
any direction without essential change of its pitch and energy. Nevertheless, at the microscopic level, this seeming freedom is achieved by correlated tilts
of discrete magnetic moments tightly bound with atoms in the crystal.
Besides, the easier seems to be the phenomenology of the isotropic system
as compared with anisotropic one, the more complex appears its description
in terms of discrete spins, particularly when the magnetic helix is
oriented by the field in some arbitrary direction. As it has been shown
recently for the helimagnetics with $B20$ structure (MnSi-type), this
freedom of the helix rotations is achieved by extra tilts (or canting) of
the local magnetic moments, which always accompany global spiralling, but
are not described in the phenomenological theory. Moreover, the
canting still remains, when the spiralling disappears in strong magnetic
field.

As a rule, in papers discussing the spin canting some class of canted
antiferromagnetics is implied \cite{cantedantifBook}, where small relative tilts of magnetic moments
belonging to different sublattices result in appearing of weak
magnetization. The phenomenon is called weak ferromagnetism, and it serves
as direct evidence of the canting, because the tilt angles of individual
spins are directly proportional to the observed spontaneous magnetization.
The canting in ferromagnetics is less elaborated, and it has been known for many decades that it is more
difficult to observe small changes of strong magnetization than small magnetization of weak ferromagnetics. Here, a refined experiment would be useful, e.g. on observation
of forbidden Bragg reflections of antiferromagnetic origin
\cite{Dmitrienko2012}.

\begin{figure}[h]
\begin{center}
\includegraphics[width=7cm]{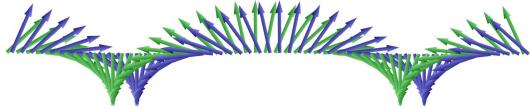}
\caption{\label{fig:canting} The twist and the canting (by example of a one-dimensional crystal with two magnetic atoms in the unit cell). The canting of spin sublattices leads to the combination of two helices with the same propagation vector and a constant phase shift.}
\end{center}
\end{figure}

Repeating small rotations (cantings) of spins belonging to
neighbouring atomic planes can result in a global twist of crystal
magnetization. Therefore, in chiral ferromagnetics, the twist and the
canting are often mistaken for the same thing. Only recently, it was
suggested how to distinguish between canting and twist in
ferromagnetics \cite{Chizhikov2012,Chizhikov2013}. It was shown that the
canting should always appear when there are more than one magnetic atom in
the unit cell. It leads to coexistence of several spin helices within the
helimagnet, which differ by phases and rotational planes (figure~\ref{fig:canting}). It was shown by example of cubic helimagnets
with $B20$ structure that the canting makes contribution to the magnetic
energy comparable with that of global spiralling. Therefore, it should be
taken into account when calculating propagation number of spin helices
from the first principles, i.e. starting from the Heisenberg microscopical
model of ferromagnetism. Even more intriguing is the fact that, if the
Dzyaloshinskii--Moriya (DM) vectors are perpendicular to the bonds between
magnetic atoms (which is predicted by many quantum mechanical models),
both twist and canting in MnSi-type crystals are determined by the same
components of the DM vectors. The study of the Cu$_2$OSeO$_3$ crystal,
with its structure different from $B20$, also can help us to distinguish
between general regularities of twisted cubic magnetics and
particularities of the MnSi-type crystals.

In this paper, a general theory of the spin canting in cubic helimagnetics
is developed, which unifies the cases of the MnSi-type and Cu$_2$OSeO$_3$
crystals. A microscopical description of the spin structures is made for the case of weakly non-collinear spins. The latter means that the angle between neighbouring
spins is close to 0 or to $\pi$, i.e. both the spiralling and canting are
small. It is just the case of the majority of itinerant magnetics with
$B20$ structure (exemplified by MnSi) and twisted ferrimagnetic
Cu$_2$OSeO$_3$. We confine ourselves only to the case of the twist induced by
the DM interactions; the case of the twist owing to the frustrated
exchange is beyond the scope of our consideration. In
section~\ref{sec:balance}, the definition of canting is given. Then, the
transition to continuous approximation is performed
(section~\ref{sec:continuous}), and the density of the magnetic energy is
calculated (section~\ref{sec:energy}). In section~\ref{sec:collinear}, it is
shown which simplifications can be made in the case, when, in the absence
of the DM interaction, all the spins of the system remain collinear. We
assert that two different kinds of the canting exist in the twisted
magnetics. The first of them is connected with the gradients of the
magnetization, and it arises even in the absence of the DM interaction (in this case the magnetization gradients can be induced by an external influence, e.g. from a surface). In
order to take this kind of canting into account, it is convenient to use
the method of fictitious ``exchange'' coordinates
(section~\ref{sec:collinear}). Then, simple expressions can be found for
the macroscopic constants ${\cal J}$ and ${\cal D}$, and the helix
propagation number $k = {\cal D} / 2 {\cal J}$
(section~\ref{sec:phenomenological}).  The second kind of canting is induced
by the DM interactions. It can be described in terms of the ``tilt''
vectors, whose symmetry is found to be determined by the symmetry of
the crystal (section~\ref{sec:collinear}). The canting of this kind still
remains even in the ferromagnetic state, when the helical spin structure
is unwound by the external magnetic field. This residual canting can be
measured, using neutron and x-ray diffractions (see discussion in
section~\ref{sec:discussion}). In sections~\ref{sec:MnSi} and
\ref{sec:Cu2OSeO3}, the theory is applied to the cases of MnSi and
Cu$_2$OSeO$_3$, respectively.

\section{Balance of magnetic moments and canting}
\label{sec:balance}

In the classical Heisenberg theory of magnetics, the energy of a
magnetic structure is written as
\begin{equation}
\label{eq:E1}
E = \sum_i \left\{ \frac{1}{2} \sum_j \left( -J_{ij} \mathbf{s}_i \cdot \mathbf{s}_j + \mathbf{D}_{ij} \cdot [\mathbf{s}_i \times \mathbf{s}_j] \right) - \mu_\mathrm{B} g_i \mathbf{H} \cdot \mathbf{s}_i \right\} ,
\end{equation}
where $i$ enumerates all the magnetic atoms, the inner sum ($j$)
is taken over close neighbours of $i$th atom, the coefficient
$\frac12$ is needed because each bond is included twice in the
sum, $\mu_\mathrm{B} g_i \mathbf{s}_i$ is the magnetic moment of $i$th atom
($\mathbf{s}_i$ is the direction of classical spin,
$|\mathbf{s}_i|=1$), $J_{ij}$ are the isotropic exchange constants, whereas
the antisymmetrical exchange is characterized by the
Dzyaloshinskii--Moriya (DM) vectors $\mathbf{D}_{ij}$
($J_{ji}=J_{ij}$, $\mathbf{D}_{ji}=-\mathbf{D}_{ij}$),
$\mathbf{H}$ is an external magnetic field.

The equation of $i$th spin balance can be easily found copying out
the part of (\ref{eq:E1}) associated with the $i$th spin,
\begin{equation}
\label{eq:Ei}
E_i = - \left\{ \sum_j \left( J_{ij} \mathbf{s}_j + [\mathbf{D}_{ij} \times \mathbf{s}_j] \right) + \mu_\mathrm{B} g_i \mathbf{H} \right\} \cdot \mathbf{s}_i \equiv -\boldsymbol{\varepsilon}_i \cdot \mathbf{s}_i
\end{equation}
(notice the absence of $\frac12$ in comparison with (\ref{eq:E1})).
It is evident that $E_i$ is minimal, when
\begin{equation}
\label{eq:si}
\mathbf{s}_i = \frac{\boldsymbol{\varepsilon}_i}{|\boldsymbol{\varepsilon}_i|} .
\end{equation}

Being of relativistic origin, the DM interaction is
considerably weaker than the isotropic exchange,
\begin{equation}
\label{eq:D/J} |\mathbf{D}_{ij}| / J_{ij} \ll 1 .
\end{equation}
Therefore it is convenient to develop a perturbation theory
with a small parameter of the order of $D/J$. In the case of
twisted magnetics we can also neglect the influence of the
external magnetic field on canting, supposing that
\begin{equation}
\label{eq:H}
H \leq H_{\mathrm{c}2} \sim \left(\frac{D}{J}\right)^2 \frac{J}{\mu_\mathrm{B}} ,
\end{equation}
where $H_{\mathrm{c}2}$ is the field of full ``unwinding'' of the magnetic
helix. Indeed, it has been found in \cite{Chizhikov2012} that the energy of canting is of the order of $(D/J)^2$, while the energy of the interaction of canting with an external magnetic field ($H<H_{\mathrm{c}2}$) is determined by the canting of the second order and being proportional to $(D/J)^4$. It follows from the finding of \cite{Dmitrienko2012} that, in ferromagnetics, the canting of the first order has antiferromagnetic character and does not contribute in this part of the magnetic energy. Notice that, accordingly to the elementary phenomenological theory, the field does not affect the helix pitch, which is in a good agreement with experimental data.

Neglecting the DM interaction and external magnetic field, equation (\ref{eq:si}) determines an
equilibrium system of spins
\begin{equation}
\label{eq:mui}
\boldsymbol{\mu}_i \equiv \mathbf{s}^{(0)}_i = \frac{\boldsymbol{\varepsilon}^{(0)}_i}{|\boldsymbol{\varepsilon}^{(0)}_i|} ,
\end{equation}
with
\begin{equation}
\label{eq:eps0}
\boldsymbol{\varepsilon}^{(0)}_i = |\boldsymbol{\varepsilon}^{(0)}_i| \boldsymbol{\mu}_i = \sum_j J_{ij} \boldsymbol{\mu}_j .
\end{equation}

\begin{figure}[h]
\begin{center}
\includegraphics[width=7cm]{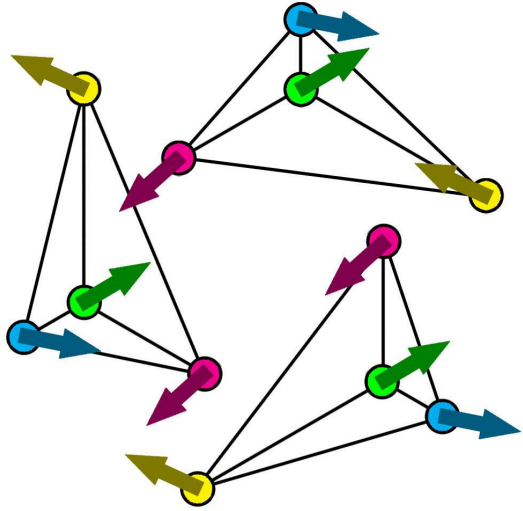}
\caption{\label{fig:equiv} Three magnetic atoms (green) in positions connected by rotational symmetry elements, and their nearest magnetic environments. In absence of non-symmetric DM interaction, a balance is achieved, when the spins in the equivalent positions (equally coloured circles) have the same directions.}
\end{center}
\end{figure}

The isotropy of the exchange interaction results in that both the
structure energy and spin balance remain unchanged, if all the
spins rotate as a unit by arbitrary angle relative to immovable
atomic structure, or, which is practically the same, if the atomic
structure rotates preserving initial spin directions. The latter
understanding is important for a cubic crystal structure, where
equivalent atomic positions connect to each other by rotational
symmetry transformations. It is obvious that a spin balance can be
achieved, when $\boldsymbol{\mu}_i$ are equal for all the
equivalent positions (figure~\ref{fig:equiv}). Here we suppose that the twist is induced only by the DM interactions, and leave the possibility of the pure exchange spiralling outside the scope of our consideration. Notice that such pure exchange twist reveals a threshold behaviour, i.e. it can appear only at some values of the exchange constants.

\begin{figure}[h]
\begin{center}
\includegraphics[width=7cm]{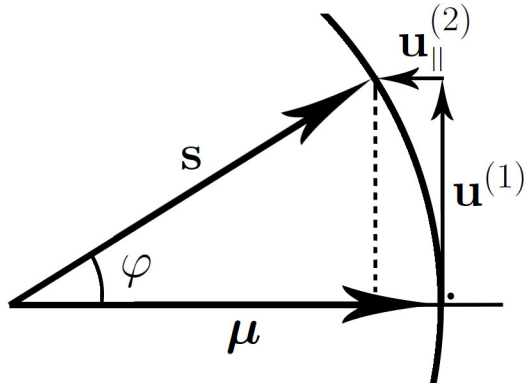}
\caption{\label{fig:circle} Deviation of spin $\mathbf{s}$ from the equilibrium state $\boldsymbol{\mu}$ in presence of non-isotropic DM interaction. The tilt angle $\varphi \sim D/J$. In a simple approximation the deviation can be decomposed into two parts, $\mathbf{u}^{(1)} \perp \boldsymbol{\mu}$ and $\mathbf{u}^{(2)}_\parallel \parallel \boldsymbol{\mu}$. }
\end{center}
\end{figure}

The non-isotropic DM interaction breaks the symmetry of $\boldsymbol{\mu}_i$,
because DM vectors of equivalent bonds change (rotate) with the bond
directions. Therefore spin $\mathbf{s}_i$ deviates from
$\boldsymbol{\mu}_i$ by a small angle of order of $D/J$. Taking
into account that $|\boldsymbol{\mu}_i|=1$, this spin change,
hereinafter referred to as ``canting'', in the first approximation
is perpendicular to $\boldsymbol{\mu}_i$. Introducing correction
$\boldsymbol{\varepsilon}^{(1)}_i$ of first order on $D/J$, we
find
\begin{equation}
\label{eq:si2}
\mathbf{s}_i \approx \frac{\boldsymbol{\varepsilon}^{(0)}_i + \boldsymbol{\varepsilon}^{(1)}_{i\parallel} + \boldsymbol{\varepsilon}^{(1)}_{i\perp}}{|\boldsymbol{\varepsilon}^{(0)}_i + \boldsymbol{\varepsilon}^{(1)}_{i\parallel} + \boldsymbol{\varepsilon}^{(1)}_{i\perp}|} \approx \boldsymbol{\mu}_i + \mathbf{u}^{(1)}_i - \frac{|\mathbf{u}^{(1)}_i|^2}{2} \boldsymbol{\mu}_i ,
\end{equation}
where $\boldsymbol{\varepsilon}^{(1)}_{i\parallel}$ and $\boldsymbol{\varepsilon}^{(1)}_{i\perp}$ are parallel and perpendicular to $\boldsymbol{\mu}_i$ components of $\boldsymbol{\varepsilon}^{(1)}_i$, correspondingly; $\mathbf{u}^{(1)}_i \equiv \boldsymbol{\varepsilon}^{(1)}_{i\perp} / |\boldsymbol{\varepsilon}^{(0)}_i|$ is the canting of $i$th spin in the first approximation, and
\begin{equation}
\label{eq:u2par}
- \frac{|\mathbf{u}^{(1)}_i|^2}{2} \boldsymbol{\mu}_i \equiv \mathbf{u}^{(2)}_{i\parallel}
\end{equation}
is proportional to $(D/J)^2$ change of the spin along
$\boldsymbol{\mu}_i$ (figure~\ref{fig:circle}).

\section{Continuous spin functions}
\label{sec:continuous}

In cubic crystal without centre of inversion, a constant direction
of magnetic moment density vector is energetically
unfavourable, and magnetic structure becomes twisted. Besides,
the strength of the twist is characterized by the same ratio $D/J$
as the canting does. If the magnetic moment density changes slowly
along the crystal, a transition can be performed from discrete
spins to smooth continuous spin functions
$\hat{\mathbf{s}}_i(\mathbf{r})$, whose number coincides with the
number of magnetic atoms in the crystal unit cell. Then, the
magnetic energy can be rewritten in integral form
\cite{Chizhikov2013},
\begin{eqnarray}
\label{eq:E2}
\fl E = \int d\mathbf{r} \sum_i \left\{ \frac{1}{2} \sum_j \left( -J_{ij} \hat{\mathbf{s}}_i \cdot \exp(\mathbf{b}_{ij} \cdot \boldsymbol{\nabla}) \hat{\mathbf{s}}_j \right. \right. \nonumber \\
\left. \vphantom{ \frac{1}{2} \sum_j} \left.  + \mathbf{D}_{ij} \cdot [\hat{\mathbf{s}}_i \times \exp(\mathbf{b}_{ij} \cdot \boldsymbol{\nabla}) \hat{\mathbf{s}}_j] \right) - \mu_\mathrm{B} g_i \mathbf{H} \cdot \hat{\mathbf{s}}_i \right\} ,
\end{eqnarray}
where the sum ($i$) is now taken over all the magnetic atoms in
the unit cell, and $\mathbf{b}_{ij}$ is a bond directed from
$i$th to $j$th atom; both $\mathbf{r}$ and $\mathbf{b}_{ij}$ are measured in cubic cell parameters $a=1$;
the values of the spin functions and their spatial derivatives in
the integrand are taken at the same point.

Using (\ref{eq:Ei}), the vector
$\hat{\boldsymbol{\varepsilon}}^{(1)}_i(\mathbf{r})$ can be
written in continuous approximation,
\begin{equation}
\label{eq:eps1}
\hat{\boldsymbol{\varepsilon}}^{(1)}_i = \sum_j \left( J_{ij} \hat{\mathbf{u}}^{(1)}_j + J_{ij} (\mathbf{b}_{ij} \cdot \boldsymbol{\nabla}) \hat{\boldsymbol{\mu}}_j + [\mathbf{D}_{ij} \times \hat{\boldsymbol{\mu}}_j] \right) ,
\end{equation}
where we neglect the magnetic field and replace
$\mathbf{u}^{(1)}_j$ and $\boldsymbol{\mu}_j$ with continuous
functions $\hat{\mathbf{u}}^{(1)}_j(\mathbf{r})$ and
$\hat{\boldsymbol{\mu}}_j(\mathbf{r})$. Here, we use condition (\ref{eq:D/J}) and make the expansion, assuming that not only canting in the first approximation but also the global gradients of the magnetic structure (e.g. the helix propagation number) are determined by small parameter $D/J$, i.e. $\boldsymbol{\nabla} \sim k \sim D/J$. Then,
\begin{equation}
\label{eq:u1}
|\hat{\boldsymbol{\varepsilon}}^{(0)}_i| \hat{\mathbf{u}}^{(1)}_i = \left\{ \sum_j \left( J_{ij} \hat{\mathbf{u}}^{(1)}_j + J_{ij} (\mathbf{b}_{ij} \cdot \boldsymbol{\nabla}) \hat{\boldsymbol{\mu}}_j + [\mathbf{D}_{ij} \times \hat{\boldsymbol{\mu}}_j] \right) \right\}_\perp ,
\end{equation}
where the subscript ``$\perp$'' designates vector projection on
the plane perpendicular to $\hat{\boldsymbol{\mu}}_i$. Because
$\hat{\mathbf{u}}^{(1)}_j \perp \hat{\boldsymbol{\mu}}_j$ and
$|\hat{\boldsymbol{\mu}}_j|=1$, then each summand in the sum over
$j$ is a vector perpendicular to $\hat{\boldsymbol{\mu}}_j$.
Equation (\ref{eq:u1}) shows that the canting is determined by both the
DM interaction and spatial derivatives of
$\hat{\boldsymbol{\mu}}$.

Let us average out (\ref{eq:u1}) using the crystal
symmetry. For any cubic (or tetrahedral) point group
\begin{equation}
\label{eq:meanbD}
\langle\mathbf{b}_{ij}\rangle_\mathrm{eq} = \langle\mathbf{D}_{ij}\rangle_\mathrm{eq} = 0 ,
\end{equation}
where index ``eq'' means averaging over equivalent bonds and positions. Thus, we obtain
\begin{equation}
\label{eq:meanu}
|\hat{\boldsymbol{\varepsilon}}^{(0)}_i| \langle\hat{\mathbf{u}}^{(1)}_i\rangle_\mathrm{eq} = \sum_j J_{ij} \langle\hat{\mathbf{u}}^{(1)}_j\rangle_{\mathrm{eq},\perp} .
\end{equation}
It is evident from comparison with (\ref{eq:eps0}), that condition (\ref{eq:meanu}) is satisfied, when
\begin{equation}
\label{eq:meanu2}
\langle\hat{\mathbf{u}}^{(1)}_i\rangle_\mathrm{eq} = [\boldsymbol{\varphi} \times \hat{\boldsymbol{\mu}}_i] ,
\end{equation}
with $[\boldsymbol{\varphi} \times \hat{\boldsymbol{\mu}}_i]$
being the change of spin $\hat{\boldsymbol{\mu}}_i$, induced by
rotation of all the spins as a unit by a small angle
$\boldsymbol{\varphi}$. As is mentioned above, the rotation
does not change the isotropic exchange energy and results in
new functions $\hat{\boldsymbol{\mu}}_i$, also satisfying
(\ref{eq:eps0}). Therefore it is possible to include
$\langle\hat{\mathbf{u}}^{(1)}_i\rangle_\mathrm{eq}$ into the definition
of $\hat{\boldsymbol{\mu}}_i$ and assume without loss of
generality that
\begin{equation}
\label{eq:meanu3}
\langle\hat{\mathbf{u}}^{(1)}_i\rangle_\mathrm{eq} = 0 .
\end{equation}
In this case, $\hat{\boldsymbol{\mu}}_i$ is the  normalized average spin over all the positions in the unit cell equivalent to the $i$th one.

\section{Energy of twist and canting}
\label{sec:energy}

Let us write out the contributions to the energy (\ref{eq:E2}) up
to the second order terms on $D/J$. The zero order energy is
\begin{equation}
\label{eq:en0}
{\cal E}^{(0)} = -\frac12 \sum_i \sum_j J_{ij} \hat{\boldsymbol{\mu}}_i \cdot \hat{\boldsymbol{\mu}}_j .
\end{equation}

The 1st order contribution
\begin{eqnarray}
\label{eq:en1}
\fl {\cal E}^{(1)} = \frac12 \sum_i \sum_j \left\{ -J_{ij} \left( \hat{\boldsymbol{\mu}}_i \cdot (\mathbf{b}_{ij} \cdot \boldsymbol{\nabla}) \hat{\boldsymbol{\mu}}_j  \vphantom{\hat{\mathbf{u}}^{(1)}_i} \right. \right. \nonumber \\
\left. \left. + \hat{\mathbf{u}}^{(1)}_i \cdot \hat{\boldsymbol{\mu}}_j + \hat{\boldsymbol{\mu}}_i \cdot \hat{\mathbf{u}}^{(1)}_j \right) + \mathbf{D}_{ij} \cdot [\hat{\boldsymbol{\mu}}_i \times \hat{\boldsymbol{\mu}}_j] \right\}
\end{eqnarray}
becomes zero in the equilibrium due to (\ref{eq:meanbD}) and (\ref{eq:meanu3}).

It is convenient to divide the 2nd order contribution into two
parts: the first one, associated with the gradients of the magnetic
moment $\hat{\boldsymbol{\mu}}$ only,
\begin{eqnarray}
\label{eq:en2mu1}
\fl {\cal E}^{(2)}_{\mu} = \frac12 \sum_i \sum_j \left\{ -\frac12 J_{ij} \hat{\boldsymbol{\mu}}_i \cdot (\mathbf{b}_{ij} \cdot \boldsymbol{\nabla})^2 \hat{\boldsymbol{\mu}}_j \right. \nonumber \\
\left. \vphantom{\frac12} + \mathbf{D}_{ij} \cdot [\hat{\boldsymbol{\mu}}_i \times (\mathbf{b}_{ij} \cdot \boldsymbol{\nabla}) \hat{\boldsymbol{\mu}}_j] \right\} ,
\end{eqnarray}
and the second one, depending on local cantings,
\begin{eqnarray}
\label{eq:en2u1}
\fl {\cal E}^{(2)}_u = \frac12 \sum_i \sum_j \left\{ -J_{ij} \left( \hat{\mathbf{u}}^{(2)}_i \cdot \hat{\boldsymbol{\mu}}_j + \hat{\boldsymbol{\mu}}_i \cdot \hat{\mathbf{u}}^{(2)}_j \right. \right. \nonumber \\
+ \hat{\mathbf{u}}^{(1)}_i \cdot (\mathbf{b}_{ij} \cdot \boldsymbol{\nabla}) \hat{\boldsymbol{\mu}}_j - \hat{\mathbf{u}}^{(1)}_j \cdot (\mathbf{b}_{ij} \cdot \boldsymbol{\nabla}) \hat{\boldsymbol{\mu}}_i \nonumber \\
\left. \left. \vphantom{\hat{\mathbf{u}}^{(2)}_j} + \hat{\mathbf{u}}^{(1)}_i \cdot \hat{\mathbf{u}}^{(1)}_j \right) + \mathbf{D}_{ij} \cdot \left( [\hat{\mathbf{u}}^{(1)}_i \times \hat{\boldsymbol{\mu}}_j] + [\hat{\boldsymbol{\mu}}_i \times \hat{\mathbf{u}}^{(1)}_j] \right) \right\} .
\end{eqnarray}

The expression for ${\cal E}^{(2)}_\mu$ can be averaged over equivalent bonds, using equation
\begin{equation}
\label{eq:meanab}
\langle a_\alpha b_\beta \rangle_\mathrm{eq} = \frac13 \delta_{\alpha\beta} (\mathbf{a} \cdot \mathbf{b})
\end{equation}
for the vectors $\mathbf{a}$ and $\mathbf{b}$ transformed with the elements of the cubic (tetrahedral) point group. Then
\begin{eqnarray}
\label{eq:en2mu-2}
\fl {\cal E}^{(2)}_{\mu} = -\frac16 \sum_i \sum_j \left\{ \frac12 J_{ij} |\mathbf{b}_{ij}|^2 \hat{\boldsymbol{\mu}}_i \cdot \Delta \hat{\boldsymbol{\mu}}_j \right. \nonumber \\
\left. \vphantom{\frac12} + (\mathbf{D}_{ij} \cdot \mathbf{b}_{ij}) \hat{\boldsymbol{\mu}}_i \cdot [\boldsymbol{\nabla} \times \hat{\boldsymbol{\mu}}_j] \right\} .
\end{eqnarray}
Notice that all the summands corresponding to equivalent bonds are
equal, and, therefore, we can calculate them only once,
multiplying then by the bond multiplicities.

Changing summation order and using relations $\mathbf{b}_{ji}=-\mathbf{b}_{ij}$ and $\mathbf{D}_{ji}=-\mathbf{D}_{ij}$, equation (\ref{eq:en2u1}) can be rewritten as
\begin{eqnarray}
\label{eq:en2u2}
\fl {\cal E}^{(2)}_u = \sum_i \sum_j \left\{ -J_{ij} \left( \hat{\mathbf{u}}^{(2)}_i \cdot \hat{\boldsymbol{\mu}}_j + \hat{\mathbf{u}}^{(1)}_i \cdot (\mathbf{b}_{ij} \cdot \boldsymbol{\nabla}) \hat{\boldsymbol{\mu}}_j \vphantom{\frac12} \right. \right. \nonumber \\
\left. \left. + \frac12 \hat{\mathbf{u}}^{(1)}_i \cdot \hat{\mathbf{u}}^{(1)}_j \right) + \mathbf{D}_{ij} \cdot [\hat{\mathbf{u}}^{(1)}_i \times \hat{\boldsymbol{\mu}}_j] \right\} .
\end{eqnarray}
Using (\ref{eq:eps0}), (\ref{eq:u2par}) and (\ref{eq:u1}), the latter expression can be transformed to
\begin{equation}
\label{eq:en2u3}
{\cal E}^{(2)}_u = -\frac12 \sum_i |\hat{\boldsymbol{\varepsilon}}^{(0)}_i| |\hat{\mathbf{u}}^{(1)}_i|^2 + \frac12 \sum_i \sum_j J_{ij} \hat{\mathbf{u}}^{(1)}_i \cdot \hat{\mathbf{u}}^{(1)}_j .
\end{equation}

\section{The case of collinear mean spins}
\label{sec:collinear}

In frustrated magnetic structures, the spins
$\hat{\boldsymbol{\mu}}_i$ can be non-collinear. Besides, in a
number of practically important cases, including ferro-,
antiferro-, and some ferrimagnetic orders, all
$\hat{\boldsymbol{\mu}}_i$ can be aligned along a line, and
several considerable simplifications can be made. Indeed, assume
that
\begin{equation}
\label{eq:muimu}
\hat{\boldsymbol{\mu}}_i = c_i \hat{\boldsymbol{\mu}} \qquad c_i = \pm 1 \qquad |\hat{\boldsymbol{\mu}}| = 1.
\end{equation}
Here, $\hat{\boldsymbol{\mu}}$ is a unit vector directed along the local magnetization. Then all $\hat{\mathbf{u}}^{(1)}_i$ belong to the plane perpendicular to $\hat{\boldsymbol{\mu}}$, and we can discard the index ``$\perp$'' in (\ref{eq:u1}):
\begin{eqnarray}
\label{eq:u1-2}
\fl \left| \sum_j J_{ij} c_j \right| \hat{\mathbf{u}}^{(1)}_i = \sum_j \left\{ J_{ij} \hat{\mathbf{u}}^{(1)}_j + J_{ij} c_j (\mathbf{b}_{ij} \cdot \boldsymbol{\nabla}) \hat{\boldsymbol{\mu}} \right. \nonumber \\
\left. \vphantom{\hat{\mathbf{u}}^{(1)}_j} + c_j [\mathbf{D}_{ij} \times \hat{\boldsymbol{\mu}}] \right\} ,
\end{eqnarray}
where is used that
\begin{equation}
\label{eq:eps0-2}
|\hat{\boldsymbol{\varepsilon}}^{(0)}_i| = \left| \sum_j J_{ij} c_j \right| .
\end{equation}

Another simplification is the use of condition
\begin{equation}
\label{eq:xideal}
\sum_j J_{ij} c_j \mathbf{b}_{ij} = 0 .
\end{equation}
The system (\ref{eq:xideal}) contains several independent vector equations (whose number coincides with the number of non-equivalent magnetic positions in the unit cell) and can be considered as a condition imposed on the atomic coordinates. Indeed, the energy (\ref{eq:E1}) does not depend explicitly on the atomic coordinates, as opposed to expression (\ref{eq:E2}), where they are included in vectors $\mathbf{b}_{ij}$. Obviously, the coordinates should disappear after minimization of the energy, therefore we can choose them arbitrarily, e.g. using condition (\ref{eq:xideal}). It is shown in \cite{Chizhikov2013}, that the possibility of intentional choice of ideal atomic coordinates is connected with the ambiguity of transition from discrete spins to continuous density of magnetic moment. Notice that, because (\ref{eq:xideal}) contains only exchange interaction parameters, these fictitious coordinates depend only on $J_{ij}$. So, hereinafter we will refer to the positions as ``exchange'' ones. In section~\ref{sec:discussion} the sense and properties of the exchange coordinates are discussed in details.

The atomic ``shift'' from real to fictitious positions results in disappearing of the part of the canting induced by derivatives of $\hat{\boldsymbol{\mu}}$:
\begin{equation}
\label{eq:u1-3}
\left| \sum_j J_{ij} c_j \right| \hat{\mathbf{u}}^{(1)}_i = \sum_j \left( J_{ij} \hat{\mathbf{u}}^{(1)}_j + c_j [\mathbf{D}_{ij} \times \hat{\boldsymbol{\mu}}] \right) .
\end{equation}

Now the canting can be found in form
\begin{equation}
\label{eq:u1rho}
\hat{\mathbf{u}}^{(1)}_i = c_i [\boldsymbol{\rho}_i \times \hat{\boldsymbol{\mu}}] ,
\end{equation}
where ``tilt'' vectors $\boldsymbol{\rho}_i$ possess the following properties. (i) The vector $\boldsymbol{\rho}_i$ corresponds to the symmetry of the $i$th atomic position, e.g. the tilt vector of an atom in position $4a$ of the space group $P2_13$ is directed along 3-fold axis of symmetry passing through the atom, whereas the tilt vector of an atom in position $12b$ has three independent components. (ii) The tilt vectors in equivalent positions are connected to each other by the corresponding transformation of the point group.

Now the system (\ref{eq:u1-3}) can be rewritten as
\begin{equation}
\label{eq:rho}
\left| \sum_j J_{ij} c_j \right| c_i \boldsymbol{\rho}_i = \sum_j \left( J_{ij} c_j \boldsymbol{\rho}_j + c_j \mathbf{D}_{ij} \right) .
\end{equation}
The system (\ref{eq:rho}) contains as many independent vector equations as many non-equivalent magnetic atoms are in the unit cell.

\section{Phenomenological constants}
\label{sec:phenomenological}

The energy density expressions are also simplified in the case of collinear mean spins:
\begin{equation}
\label{eq:en0-2}
{\cal E}^{(0)} = -\frac12 \sum_i \sum_j J_{ij} c_i c_j ,
\end{equation}
\begin{eqnarray}
\label{eq:en2mu-3}
\fl {\cal E}^{(2)}_{\mu} = -\frac{1}{12} \left( \sum_i \sum_j J_{ij} c_i c_j |\mathbf{b}_{ij}|^2 \right) \hat{\boldsymbol{\mu}} \cdot \Delta \hat{\boldsymbol{\mu}} \nonumber \\
- \frac16 \left( \sum_i \sum_j c_i c_j \mathbf{D}_{ij} \cdot \mathbf{b}_{ij} \right) \hat{\boldsymbol{\mu}} \cdot [\boldsymbol{\nabla} \times \hat{\boldsymbol{\mu}}] .
\end{eqnarray}

Using equation
\begin{equation}
\label{eq:rhorho-3}
\langle [\boldsymbol{\rho}_i \times \hat{\boldsymbol{\mu}}] \cdot [\boldsymbol{\rho}_j \times \hat{\boldsymbol{\mu}}] \rangle_\mathrm{eq} = \frac23 \boldsymbol{\rho}_i \cdot \boldsymbol{\rho}_j ,
\end{equation}
we find
\begin{equation}
\label{eq:en2u4}
{\cal E}^{(2)}_u = -\frac13 \sum_i \left| \sum_j J_{ij} c_j \right| |\boldsymbol{\rho}_i|^2 + \frac13 \sum_i \sum_j J_{ij} c_i c_j \boldsymbol{\rho}_i \cdot \boldsymbol{\rho}_j
\end{equation}
or, using (\ref{eq:rho}),
\begin{equation}
\label{eq:en2u5}
{\cal E}^{(2)}_u = -\frac13 \sum_i \sum_j c_i c_j \mathbf{D}_{ij} \cdot \boldsymbol{\rho}_i .
\end{equation}
Notice that in the latter expression as well as in (\ref{eq:en2mu-3}) the summands corresponding to equivalent bonds are equal. In the given approximation, expression (\ref{eq:en2u5}) is nothing but a constant (negative) addition to the magnetic energy. However, it was found in \cite{Chizhikov2013} that this contribution had view $f(M) \sim \sqrt{M_0^2-M^2}$, with $M_0$ being a saturation magnetization. This term is important for describing of double twisted magnetic structures, e.g. the Skyrmion lattices (A-phases) and hypothetical cubic textures similar to the blue phases of liquid crystals. Indeed, it is not clear so far why the A-phases are energetically preferable in some ranges of temperatures and magnetic fields. In order to solve this problem, in the phenomenological theory, they often add to the energy a correction term in the form $f(M) \sim aM^2 + bM^4$. Yet a trouble emerges here, namely, the magnetization in some areas can exceed the saturation value $M_0$. The form $f(M) \sim \sqrt{M_0^2-M^2}$, resulting from the Heisenberg model, resolves this difficulty.

Because, as is seen from (\ref{eq:rho}),
$\boldsymbol{\rho}_i$ are independent of spatial derivatives
of $\hat{\boldsymbol{\mu}}$, the energy ${\cal E}^{(2)}_u$ does
not contain the derivatives as well. On the other hand, ${\cal
E}^{(2)}_\mu$ does not depend on $\boldsymbol{\rho}_i$, and this part of the
energy can be rewritten in the following form conventional for
the macroscopic phenomenological theory
\begin{equation}
\label{eq:en2mu-4}
{\cal E}^{(2)}_{\mu} = {\cal J} \frac{\partial \hat{\mu}_i}{\partial x_k} \frac{\partial \hat{\mu}_i}{\partial x_k} + {\cal D}  \hat{\boldsymbol{\mu}} \cdot [\boldsymbol{\nabla} \times \hat{\boldsymbol{\mu}}] ,
\end{equation}
where ${\cal J}$ and ${\cal D}$ are the phenomenological constants
expressed now through the parameters of the microscopical
Heisenberg model:
\begin{equation}
\label{eq:calJ}
{\cal J} = \frac{1}{12} \sum_i \sum_j J_{ij} c_i c_j |\mathbf{b}_{ij}|^2 ,
\end{equation}
\begin{equation}
\label{eq:calD}
{\cal D} = -\frac16 \sum_i \sum_j c_i c_j \mathbf{D}_{ij} \cdot \mathbf{b}_{ij} ,
\end{equation}
and $\hat{\boldsymbol{\mu}}$ is a unit vector oriented along the
local magnetic moment density $\mathbf{M}$. Parameters
${\cal J}$ and ${\cal D}$ determine spiralling of the spin
structure. In particular, the minimization of
(\ref{eq:en2mu-4}) gives as a result the helix with
propagation number
\begin{equation}
\label{eq:k}
k = \frac{\cal D}{2{\cal J}}
\end{equation}
and vector $\hat{\boldsymbol{\mu}}$ rotating in the plane perpendicular to the helix axis.

The absence of canting in expression (\ref{eq:en2mu-4}) does not mean that the canting does not affect the spiralling. As a matter of fact, the contribution of canting is taken into account, when choosing the fictitious atomic coordinates with condition (\ref{eq:xideal}). The simple additive expressions (\ref{eq:calJ}) and (\ref{eq:calD}) can appear only because the contributing bonds $\mathbf{b}_{ij}$ are the functions of the exchange coordinates. Notice that, due to the dependence of vectors $\mathbf{b}_{ij}$ on the exchange constants, both ${\cal J}$ and ${\cal D}$ are non-linear functions of $J_{ij}$.
In the case of $n$ non-equivalent bonds with parameters $J_1 \ldots J_n$, $\mathbf{D}_1 \ldots \mathbf{D}_n$ and $m$ independent coordinates of magnetic atoms ($m=1$ for MnSi and $m=4$ for Cu$_2$OSeO$_3$), the phenomenological constants have the following views: ${\cal J} = P_{2m+1}(J_1 \ldots J_n) / P_m^2(J_1 \ldots J_n)$, ${\cal D} = P_{1,m}(\mathbf{D}_1 \ldots \mathbf{D}_n, J_1 \ldots J_n) / P_m(J_1 \ldots J_n)$, where $P_{2m+1}$ and $P_m$ are some homogeneous polynomials of $J_1 \ldots J_n$ with degrees $2m+1$ and $m$, correspondingly, and $P_{1,m}$ is a homogeneous polynomial linear by the components of vectors $\mathbf{D}_1 \ldots \mathbf{D}_n$ and being of degree $m$ with respect to $J_1 \ldots J_n$. Indeed, it follows from equation (\ref{eq:xideal}), which is a system of $m$ linear equations with respect exchange coordinates, with both coefficients and constant terms being linear functions of $J_1 \ldots J_n$. Consequently, the intuitively obvious for dimensional reason expression $k \sim D/J$ transforms to $k \sim P_{1,m}(\mathbf{D}_1 \ldots \mathbf{D}_n, J_1 \ldots J_n) P_m(J_1 \ldots J_n) / P_{2m+1}(J_1 \ldots J_n)$.

\section{The example of the $\mathbf{MnSi}$-type crystals}
\label{sec:MnSi}

Let us illustrate the theory developed by a simple example of the MnSi-type crystals with the $B20$ structure. The cubic crystal MnSi has the space group $P2_13$. Its unit cell contains four magnetic manganese atoms in the position $4a$ (at threefold axes) with $x = 0.138$. In accordance with \cite{Chizhikov2013}, we define four magnetic shells by the non-equivalent bonds $\mathbf{b}_1 = (-2x, \frac12, \frac12-2x)$, $\mathbf{b}_2 = (1-2x, \frac12, \frac12-2x)$, $\mathbf{b}_3 = (-2x, \frac12, -\frac12-2x)$, and $\mathbf{b}_4 = (1, 0, 0)$, with corresponding exchange constants $J_1$--$J_4$ and DM vectors $\mathbf{D}_1$--$\mathbf{D}_4$.

The first atom $(x, x, x)$ has 6 bonds of each kind: $(-2x, \pm\frac12, \frac12-2x) \circlearrowleft$, $(1-2x, \pm\frac12, \frac12-2x) \circlearrowleft$, $(-2x, \pm\frac12, -\frac12-2x) \circlearrowleft$, $\pm(1, 0, 0) \circlearrowleft$ (the symbol $\circlearrowleft$ means possible cyclic permutations of the coordinates). The corresponding DM vectors can be obtained from $\mathbf{D}_1$--$\mathbf{D}_4$ by the same sing changes and cyclic permutations of the components.

We assign $c_i = 1$ for all the manganese atoms in the equivalent positions. Then, condition (\ref{eq:xideal}) for the atom $(x, x, x)$ can be rewritten as
\begin{equation}
\label{eq:condition}
\sum_j J_{ij} \mathbf{b}_{ij} = 0
\end{equation}
or, using the symmetry 3 of the position,
\begin{equation}
\label{eq:condition-2}
J_1 \left( \frac12-4x \right) + J_2 \left( \frac32-4x \right) + J_3 \left( -\frac12-4x \right) + J_4 \cdot 0 = 0
\end{equation}
and
\begin{equation}
\label{eq:xideal-MnSi}
x_\mathrm{exch} = \frac{J_1+3J_2-J_3}{8(J_1+J_2+J_3)}
\end{equation}
in accordance with \cite{Chizhikov2013}. The 4th neighbours do not influence the exchange coordinate, because the bond $\mathbf{b}_4$ connect the atoms belonging to the same magnetic sublattice.

From (\ref{eq:calJ}) and (\ref{eq:calD}) the macroscopic parameters can be found,
\begin{equation}
\label{eq:calJ-MnSi}
{\cal J} = 2 (J_1 b_1^2 + J_2 b_2^2 + J_3 b_3^2 + J_4 b_4^2) ,
\end{equation}
\begin{equation}
\label{eq:calD-MnSi}
{\cal D} = -4 (\mathbf{D}_1 \cdot \mathbf{b}_1 + \mathbf{D}_2 \cdot \mathbf{b}_2 + \mathbf{D}_3 \cdot \mathbf{b}_3 + \mathbf{D}_4 \cdot \mathbf{b}_4) .
\end{equation}
After substitution of the exchange coordinate $x_\mathrm{exch}$ into the $\mathbf{b}$ vectors, the macroscopic exchange parameter can be expressed through the exchange constants of the bonds,
\begin{equation}
\label{eq:calJ-2}
{\cal J} = \frac{3J_1^2+3J_2^2+3J_3^2+10J_1J_2+10J_1J_3+22J_2J_3}{4(J_1+J_2+J_3)} + 2J_4 ,
\end{equation}
which coincides with the result from \cite{Chizhikov2013}.

The atom $(x, x, x)$, belonging to the 1st magnetic sublattice of the crystal, has two atoms of each other sublattice (2, 3, 4) in the 1st, the 2nd and the 3rd magnetic coordination spheres, and six atoms of the same sublattice (1) in the 4th coordination sphere. Therefore, condition~(\ref{eq:rho}) on tilt vectors can be written for the atom $(x, x, x)$ as
\begin{eqnarray}
\label{eq:rho-MnSi}
\fl 6(J_1+J_2+J_3+J_4) \boldsymbol{\rho}_1 = 2 (J_1+J_2+J_3) (\boldsymbol{\rho}_2 + \boldsymbol{\rho}_3 + \boldsymbol{\rho}_4) \nonumber \\
+ 6J_4 \boldsymbol{\rho}_1 + 2 D_+ (1,1,1) ,
\end{eqnarray}
with $D_+ = D_{1x} + D_{1z} + D_{2x} + D_{2z} + D_{3x} + D_{3z}$. Using symmetry equation
\begin{equation}
\label{eq:sumrho}
\boldsymbol{\rho}_1 + \boldsymbol{\rho}_2 + \boldsymbol{\rho}_3 + \boldsymbol{\rho}_4 = 0 ,
\end{equation}
we find
\begin{equation}
\label{eq:rho-3}
\boldsymbol{\rho}_1 = (\rho_x, \rho_x, \rho_x) \qquad \rho_x = \frac{D_+}{4(J_1+J_2+J_3)}
\end{equation}
in accordance with \cite{Chizhikov2013}. Tilt vectors $\boldsymbol{\rho}_2$, $\boldsymbol{\rho}_3$ and $\boldsymbol{\rho}_4$ of other manganese atoms in the unit cell can be obtained from $\boldsymbol{\rho}_1$ by the corresponding symmetry transformations. As is seen from (\ref{eq:rho-3}), the 4th magnetic neighbours, belonging to the same sublattice, do not influence tilt vectors.

\section{The case of $\mathbf{Cu_2OSeO_3}$}
\label{sec:Cu2OSeO3}

\begin{figure}[h]
\begin{center}
\includegraphics[width=7cm]{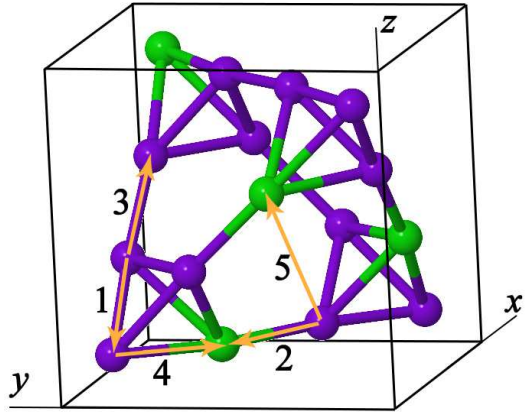}
\caption{\label{fig:unitcell} Copper atoms in the unit
cell of Cu$_2$OSeO$_3$: four Cu-I (green) in positions $4a$ and twelve
Cu-II (indigo) in positions $12b$ of the space group $P2_13$. The arrows
show five non-equivalent bonds providing main contributions into the
isotropic exchange and the DM interactions.}
\end{center}
\end{figure}

The cubic crystal Cu$_2$OSeO$_3$ has the space group $P2_13$, the
lattice constant $a=8.925$ \AA. Its unit cell contains 16 magnetic
copper atoms: four Cu-I in the position $4a$ with coordinates
$\mathbf{r}_\mathrm{I} = (x_\mathrm{I}, x_\mathrm{I}, x_\mathrm{I}) = (0.8860, 0.8860,0.8860)$ (at threefold axes) and twelve Cu-II in the general position
$12b$ with coordinates $\mathbf{r}_\mathrm{II} = (0.1335, 0.1211,
0.8719)$ \cite{Effenberger1986}. In the unit cell, 16 copper atoms form four almost regular tetrahedra, which, in their turn, do a bigger one, resembling the first-order model of the Sierpinski fractal tetrahedron (figure~\ref{fig:unitcell}). All the bonds shown by the rods in the picture have approximately the same length. The nearest magnetic environment
is determined by four types of non-equivalent bonds, but following
\cite{GongXiang2012} we take into account an additional
bond with the atoms of the second magnetic environment. The examples of these five non-equivalent bonds are represented by the arrows in figure~\ref{fig:unitcell}. In table~\ref{5bonds}, the coordinates and the energetic parameters are listed of five non-equivalent bonds, radiated from the Cu-II atom in position $\mathbf{r}_\mathrm{II}$.

\begin{table}
\caption{\label{5bonds} Cu$_2$OSeO$_3$ crystal: five non-equivalent
bonds between copper atoms with the corresponding exchange
constants $J$ and DM vectors $\mathbf{D}$, taken from
\cite{GongXiang2012}. The lattice parameter $a = 8.925 $\AA\
and atomic coordinates are as in \cite{Effenberger1986}.}
\begin{indented}
\item[]\begin{tabular}{@{}clll}
\br
$\mathbf{b}+(x_\mathrm{II},y_\mathrm{II},z_\mathrm{II})$ & $a|\mathbf{b}|$,\AA & $J$, meV & $\mathbf{D}$, meV \\
\mr
$-\frac12+z_\mathrm{II}, \frac12-x_\mathrm{II}, 1-y_\mathrm{II}$ & 3.054 & $J_1 = 1.132$ & $\mathbf{D}_1 = (0.289, -0.325, -0.051)$ \\
$-1+x_\mathrm{I}, -1+x_\mathrm{I}, x_\mathrm{I}$ & 3.049 & $J_2 = -6.534$ & $\mathbf{D}_2 = (1.12, -1.376, 0.300)$ \\
$-1+z_\mathrm{II}, x_\mathrm{II}, 1+y_\mathrm{II}$ & 3.226 & $J_3 = 3.693$ & $\mathbf{D}_3 = (-0.263, 0.167, -0.407)$ \\
$1-x_\mathrm{I}, -\frac12+x_\mathrm{I}, \frac32-x_\mathrm{I}$ & 3.304 & $J_4 = -0.900$ & $\mathbf{D}_4 = (-0.490, 1.238, 1.144)$ \\
$\frac12-x_\mathrm{I}, 1-x_\mathrm{I}, -\frac12+x_\mathrm{I}$ & 6.349 & $J_5 = -0.984$ & $\mathbf{D}_5 = (0.045, -0.087, -0.059)$ \\
\br
\end{tabular}
\end{indented}
\end{table}

The magnetic neighbours of atom Cu-I are nine copper atoms in positions $12b$. Three of them are listed in table~\ref{bondsI}. Because the atom in position $\mathbf{r}_\mathrm{I}$ is on the 3-fold axis $[111]$, the remaining six bonds can be obtained by cyclic permutations of the coordinates of the vectors.

\begin{table}
\caption{\label{bondsI} Magnetic neighbours of copper
Cu-I in position $\mathbf{r}_\mathrm{I}$, energy parameters of the bonds and tilt vectors
$\boldsymbol{\rho}$. Only three neighbours are listed. Other six
bonds can be obtained by threefold rotations, i.e. by cyclic
permutations of the coordinates of the vectors.}
\begin{indented}
\item[]\begin{tabular}{@{}cccc}
\br
$\mathbf{r}_j=\mathbf{r}_\mathrm{I}+\mathbf{b}_{ij}$ & $J_{ij}$ & $\mathbf{D}_{ij}$ & $\boldsymbol{\rho}_j$ \\
\mr
$1+x_\mathrm{II}, 1+y_\mathrm{II}, z_\mathrm{II}$ & $J_2$ & $-D_{2x}, -D_{2y}, -D_{2z}$ & $\rho_{\mathrm{II},x}, \rho_{\mathrm{II},y}, \rho_{\mathrm{II},z}$ \\
$1-x_\mathrm{II}, \frac12+y_\mathrm{II}, \frac32-z_\mathrm{II}$ & $J_4$ & $D_{4x}, -D_{4y}, D_{4z}$ & $-\rho_{\mathrm{II},x}, \rho_{\mathrm{II},y}, -\rho_{\mathrm{II},z}$ \\
$\frac12-x_\mathrm{II}, 1-y_\mathrm{II}, \frac12+z_\mathrm{II}$ & $J_5$ & $D_{5x}, D_{5y}, -D_{5z}$ & $-\rho_{\mathrm{II},x}, -\rho_{\mathrm{II},y}, \rho_{\mathrm{II},z}$ \\
\br
\end{tabular}
\end{indented}
\end{table}

The magnetic environment of atom Cu-II contains seven copper atoms: four in positions $12b$ and three in positions $4a$ (table~\ref{bondsII}).

\begin{table}
\caption{\label{bondsII} Magnetic neighbours of copper Cu-II in position $\mathbf{r}_\mathrm{II}$, energy parameters of the bonds and tilt vectors $\boldsymbol{\rho}$.}
\begin{indented}
\item[]\begin{tabular}{@{}cccc}
\br
$\mathbf{r}_j=\mathbf{r}_\mathrm{II}+\mathbf{b}_{ij}$ & $J_{ij}$ & $\mathbf{D}_{ij}$ & $\boldsymbol{\rho}_j$ \\
\mr
$-\frac12+z_\mathrm{II}, \frac12-x_\mathrm{II}, 1-y_\mathrm{II}$ & $J_1$ & $D_{1x}, D_{1y}, D_{1z}$ & $\rho_{\mathrm{II},z}, -\rho_{\mathrm{II},x}, -\rho_{\mathrm{II},y}$ \\
$\frac12-y_\mathrm{II}, 1-z_\mathrm{II}, \frac12+x_\mathrm{II}$ & $J_1$ & $D_{1y}, D_{1z}, -D_{1x}$ & $-\rho_{\mathrm{II},y}, -\rho_{\mathrm{II},z}, \rho_{\mathrm{II},x}$ \\
$-1+x_\mathrm{I}, -1+x_\mathrm{I}, x_\mathrm{I}$ & $J_2$ & $D_{2x}, D_{2y}, D_{2z}$ & $\rho_{\mathrm{I},x}, \rho_{\mathrm{I},x}, \rho_{\mathrm{I},x}$ \\
$-1+z_\mathrm{II}, x_\mathrm{II}, 1+y_\mathrm{II}$ & $J_3$ & $D_{3x}, D_{3y}, D_{3z}$  & $\rho_{\mathrm{II},z}, \rho_{\mathrm{II},x}, \rho_{\mathrm{II},y}$ \\
$y_\mathrm{II}, -1+z_\mathrm{II}, 1+x_\mathrm{II}$ & $J_3$ & $-D_{3y}, -D_{3z}, -D_{3x}$  & $\rho_{\mathrm{II},y}, \rho_{\mathrm{II},z}, \rho_{\mathrm{II},x}$ \\
$1-x_\mathrm{I}, -\frac12+x_\mathrm{I}, \frac32-x_\mathrm{I}$ & $J_4$ & $D_{4x}, D_{4y}, D_{4z}$ & $-\rho_{\mathrm{I},x}, \rho_{\mathrm{I},x}, -\rho_{\mathrm{I},x}$ \\
$\frac12-x_\mathrm{I}, 1-x_\mathrm{I}, -\frac12+x_\mathrm{I}$ & $J_5$ & $D_{5x}, D_{5y}, D_{5z}$ & $-\rho_{\mathrm{I},x}, -\rho_{\mathrm{I},x}, \rho_{\mathrm{I},x}$ \\
\br
\end{tabular}
\end{indented}
\end{table}

All 16 copper positions in the unit cell are characterized by
``sense'' numbers $c_i$ and tilt vectors
$\boldsymbol{\rho}_i$ determining canting. Owing to ferrimagnetic
order, the spins of the copper atoms in positions $4a$ are
opposite to those in positions $12b$ and to the summary magnetic
moment \cite{Bos2008}, so we can assign the values $c_\mathrm{I}=-1$ and
$c_\mathrm{II}=1$. The tilt vectors of the atoms in positions $\mathbf{r}_\mathrm{I}$ and $\mathbf{r}_\mathrm{II}$ are chosen as $(\rho_{\mathrm{I},x}, \rho_{\mathrm{I},x},
\rho_{\mathrm{I},x})$ and $(\rho_{\mathrm{II},x}, \rho_{\mathrm{II},y}, \rho_{\mathrm{II},z})$,
correspondingly. Tilt vectors of other copper atoms can be
obtained from these two vectors using symmetry transformations of
the point group 23.

Let us use condition~(\ref{eq:xideal}) for the atoms
in positions $\mathbf{r}_\mathrm{I}$ and $\mathbf{r}_\mathrm{II}$, in order to find
exchange coordinates of the positions. Then, the system of
linear equations can be easily obtained and solved:
\begin{equation}
\label{eq:system-coord}
{\cal A} \left( \begin{array}{c} x_\mathrm{I} \\ x_\mathrm{II} \\ y_\mathrm{II} \\ z_\mathrm{II} \end{array} \right) = \left( \begin{array}{c}
-2J_2 -3J_4 -2J_5 \\
-J_2 +J_3 +J_4 +\frac12 J_5 \\
-\frac32 J_1 -J_2 +J_3 -\frac12 J_4 +J_5 \\
-\frac32 J_1 -2J_3 +\frac32 J_4 -\frac12 J_5
\end{array} \right) ,
\end{equation}
with
\begin{equation}
\label{eq:calA}
{\cal A} = \left( \begin{array}{rccc}
A_{11} & A_{12} & A_{13} & A_{14} \\
-A_{12} & A_{22} & A_{23} & A_{24} \\
-A_{13} & A_{23} & A_{22} & A_{23} \\
-A_{14} & A_{24} & A_{23} & A_{22}
\end{array} \right) \\
\end{equation}
\begin{equation}
\label{eq:calA2}
\begin{array}{l}
A_{11} = -3J_2-3J_4-3J_5 \\
A_{12} = J_2-J_4-J_5 \\
A_{13} = J_2+J_4-J_5 \\
A_{14} = J_2-J_4+J_5 \\
A_{22} = -2J_1+J_2-2J_3+J_4+J_5 \\
A_{23} = -J_1+J_3 \\
A_{24} = J_1+J_3 .
\end{array}
\end{equation}

General solution is rather combersome. Using the values of $J_1$--$J_5$ from table~\ref{5bonds}, we find
\begin{equation}
\label{eq:coord}
\begin{array}{l}
x_{\mathrm{I},\mathrm{exch}} = 0.9417 , \\
x_{\mathrm{II},\mathrm{exch}} = -0.0042 , \\
y_{\mathrm{II},\mathrm{exch}} = 0.0202 , \\
z_{\mathrm{II},\mathrm{exch}} = 0.8969 .
\end{array}
\end{equation}
Just these ideal coordinates, rather than the real ones, should be
used in calculations of the phenomenological constants ${\cal J}$,
${\cal D}$ by formulae (\ref{eq:calJ}), (\ref{eq:calD}). The
calculation with the data from table~\ref{5bonds} gives ${\cal
J}=2.565$ meV, ${\cal D}=0.970$ meV. Therefore, the helix
propagation number $k={\cal D}/2{\cal J}= 0.1890$ ($q=k/2\pi=0.0301$). The positive value of $k$ means that the magnetic helicoid
is expected to be right-handed in the crystals with the given set
of atomic positions. In the enantiomorphs with opposite values of
all atomic positions the helicoid should be also opposite
(left-handed). This prediction is rather easy for experimental
proof (the corresponding procedure is well developed for MnSi-type crystals \cite{Dmitriev2012,Tanaka1985,Ishida1985,Grigoriev2013,Shibata2013}).

In order to find the tilt vectors, we write system~(\ref{eq:rho}) for the atoms in positions $\mathbf{r}_\mathrm{I}$ and $\mathbf{r}_\mathrm{II}$,
\begin{equation}
\label{eq:system-rho2}
{\cal A} \left( \begin{array}{c} \rho_{\mathrm{I},x} \\ \rho_{\mathrm{II},x} \\ \rho_{\mathrm{II},y} \\ \rho_{\mathrm{II},z} \end{array} \right) = \left( \begin{array}{c} B_1 \\ B_2 \\ B_3 \\ B_4 \end{array} \right)
\end{equation}
\begin{equation}
\label{eq:B}
\begin{array}{ll}
B_1 = & D_{2x}+D_{2y}+D_{2z}-D_{4x}+D_{4y}-D_{4z} \\
& -D_{5x}-D_{5y}+D_{5z} \\
B_2 = & -D_{1x}-D_{1y}+D_{2x}-D_{3x}+D_{3y}+D_{4x}+D_{5x} \\
B_3 = & -D_{1y}-D_{1z}+D_{2y}-D_{3y}+D_{3z}+D_{4y}+D_{5y} \\
B_4 = & D_{1x}-D_{1z}+D_{2z}+D_{3x}-D_{3z}+D_{4z}+D_{5z} .
\end{array}
\end{equation}
Using the data from table~\ref{5bonds}, we find
\begin{equation}
\label{eq:rho2}
\begin{array}{l}
\rho_{\mathrm{I},x} = 0.049 , \\
\rho_{\mathrm{II},x} = -0.123 , \\
\rho_{\mathrm{II},y} = -0.034 , \\
\rho_{\mathrm{II},z} = -0.159 .
\end{array}
\end{equation}
The canting angles are $|\boldsymbol{\rho}_\mathrm{I}|=0.084$ ($4.8^\circ$), $|\boldsymbol{\rho}_\mathrm{II}|=0.204$ ($11.7^\circ$).

\begin{figure}[h]
\begin{center}
\includegraphics[width=7cm]{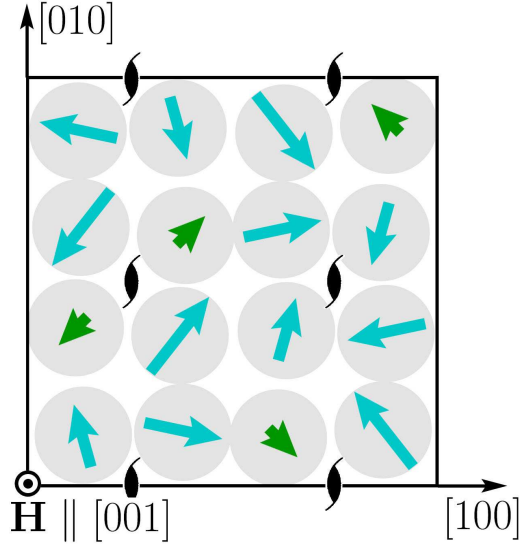}
\caption{\label{fig:ferro} The spin canting in the
ferrimagnetic state of Cu$_2$OSeO$_3$, induced by an external magnetic
field applied in the [001] direction. The centres of the grey circles
coincide with projections of copper atoms in plane (001). The
green and cyan arrows show directions and magnitudes of spin
cantings of 4 Cu-I and 12 Cu-II atoms, correspondingly.}
\end{center}
\end{figure}

The tilt vectors (\ref{eq:rho2}) determine spin cantings
$\mathbf{u}^{(1)}$ of the first approximation, both in arbitrary twisted
phases, including helicoids and A-phase, and in the state unwound by
magnetic field $H \gtrsim H_{\mathrm{c}2}$. For example, figure~\ref{fig:ferro} shows
the canting arrangement in the periodic structure, if the external
magnetic field is along the [001] axis. In the first approximation,
all cantings lie in plane (001), perpendicular to the field and
magnetization. It is obvious that the canting arrangement is
symmetrical relative to the $2_1$ screw axes directed along the field.

The absolute atomic configuration of Cu$_2$OSeO$_3$ was determined in
\cite{Effenberger1986}. We find that it is very close to $P4_132$
cubic symmetry \cite{IntTabA}, see table~\ref{tabstruc} where the
idealized structure with the $P4_132$ symmetry is compared with the
observed structure. Thus, similar to the case of the $B20$ structures
\cite{Dmitriev2012}, we can say that this atomic structure is right-handed
because the space group $P4_132$ contains only right-handed screw axes
$4_1$. And in this right-handed structure the magnetic helicoids are
expected to be right-handed according to above calculations. For another
enantiomorph with inverse values of all atomic coordinates, the idealized
atomic structure is left-handed ($P4_332$ space group). The corresponding
inverted real structure (it has exactly the same energy as the structure
determined in \cite{Effenberger1986}) can be also considered as
left-handed and its magnetic helicoid is expected to be left-handed, but
its space group remains $P2_13$.

\begin{table}
\caption{\label{tabstruc} The atomic structure of Cu$_2$OSeO$_3$
\cite{Effenberger1986} with $P2_13$ symmetry and the idealized structure
with $P4_132$ symmetry.}
\begin{indented}
\item[]\begin{tabular}{@{}lllll}
\br
\multirow{2}{*}{atoms} & \multicolumn{2}{l}{$P2_13$, \cite{Effenberger1986}} & \multicolumn{2}{l}{$P4_132$} \\
\cline{2-5}
 & type & $x,y,z$ & type & $x,y,z$ \\
\mr
Cu-I & $4a$ & 0.8860,0.8860,0.8860 & $4b$ & $\frac78,\frac78,\frac78$ \\
Cu-II & $12b$ & 0.1335,0.1211,-0.1281 & $12d$ & $y+\frac14,\frac18,y \quad (y \approx -0.12)$ \\
Se-I & $4a$ & 0.4590,0.4590,0.4590 & \multirow{2}{*}{$\left. \vphantom{\frac12} \right\}8c$} & \multirow{2}{*}{$x,x,x \quad (x \approx 0.5)$} \\
Se-II & $4a$ & 0.2113,0.2113,0.2113 & & \\
O-I & $4a$ & 0.0105,0.0105,0.0105 & \multirow{2}{*}{$\left. \vphantom{\frac12} \right\}8c$} & \multirow{2}{*}{$x,x,x \quad (x \approx 0.0)$} \\
O-II & $4a$ & 0.7621,0.7621,0.7621 & & \\
O-III & $12b$ & 0.2699,0.4834,0.4706 & $12d$ & $-y,y+\frac34,\frac38 \quad (y \approx -0.27)$ \\
O-IV & $12b$ & 0.2710,0.1892,0.0313 & $12d$ & $y+\frac14,\frac18,y \quad (y \approx 0.03)$ \\
\br
\end{tabular}
\end{indented}
\end{table}

Another geometrical approach is to reduce the magnetic structure of Cu$_2$OSeO$_3$ to that of the MnSi-type crystals \cite{JansonDMI2013}. For that, the structure is divided into ``strong'' tetrahedra of copper atoms, with the summary magnetic moments being considered as individual classical spins. Indeed, as is seen from table~\ref{5bonds}, the bonds 2 and 3 have the maximal absolute values of the exchange interaction constants. The approximation of strong tetrahedra corresponds to the infinite constants: $J_2 = -\infty$, $J_3 = +\infty$. Then, the corresponding bonds with the exchange coordinates become zero, $\mathbf{b}_2 = \mathbf{b}_3 = 0$, and the tetrahedra transform to ``atoms'' in positions $4a$ with coordinate
\begin{equation}
\label{eq:toB20}
x_\mathrm{exch} \equiv x_{\mathrm{I},\mathrm{exch}} - 1 = x_{\mathrm{II},\mathrm{exch}} = y_{\mathrm{II},\mathrm{exch}} = z_{\mathrm{II},\mathrm{exch}} - 1 .
\end{equation}
Besides, the bonds $\mathbf{b}_1$ of Cu$_2$OSeO$_3$ transform into the bonds $\mathbf{b}_1$ of MnSi, whereas the bonds $\mathbf{b}_4$ and $\mathbf{b}_5$ of Cu$_2$OSeO$_3$ become the bonds $\mathbf{b}_3$ of MnSi. The coordinate $x_\mathrm{exch}$ can be found from the solution of the system~(\ref{eq:system-coord}), when passing to the limit $J_2 \rightarrow -\infty$, $J_3 \rightarrow +\infty$. But, it is easier to use expression (\ref{eq:xideal-MnSi}) for MnSi with the parameters $J_1(\mbox{MnSi}) = J_1(\mbox{Cu$_2$OSeO$_3$})$, $J_2(\mbox{MnSi}) = 0$, and $J_3(\mbox{MnSi}) = -(J_4 + J_5)(\mbox{Cu$_2$OSeO$_3$})$ (the sign minus is due to bonds $\mathbf{b}_4$ and $\mathbf{b}_5$ connecting atoms Cu-I and Cu-II having opposite spins). Therefore,
\begin{equation}
\label{eq:toB20-2}
x_\mathrm{exch} = \frac{J_1 + J_4 + J_5}{8(J_1 - J_4 - J_5)} (\mbox{Cu$_2$OSeO$_3$}) = -0.0312 .
\end{equation}
Then, from (\ref{eq:calJ})--(\ref{eq:k}) can be calculated ${\cal J}=2.969$ meV, ${\cal D}=0.534$ meV, and $k=0.0900$. As we see, the approximation of ``strong'' tetrahedra with the data from \cite{GongXiang2012} gives value of the propagation number $k$, which is two times less than one obtained with true energetic parameters of bonds $\mathbf{b}_2$ and $\mathbf{b}_3$.

\section{Discussion}
\label{sec:discussion}

The calculated value of the propagation number, $k=0.189$, is almost three times greater than one found in \cite{GongXiang2012} without taking the canting into account, and approximately two times bigger than the experimental value $k=0.088$ \cite{AdamsPRL2012,SekiPRB2012}. Let us explain the cause of it. In order to calculate $k$ in the one-helix model (without canting), the formulae (\ref{eq:calJ})--(\ref{eq:k}) can be used with real coordinates instead of exchange ones. Indeed, in this case, using the data from \cite{GongXiang2012}, we find $k=0.07$, which is very close to the value calculated in \cite{GongXiang2012} without taking the canting into account. In order to understand this essential difference, notice that, accordingly to (\ref{eq:calD}), the phenomenological parameter ${\cal
D}$ determining spiralling of the magnetic structure is composed
of the scalar products $\mathbf{D} \cdot \mathbf{b}$. It is well known from the superexchange theory, that
the DM vectors are approximately perpendicular to the bonds (see
also third column in table~\ref{bexch}). Were the perpendicularity of the DM vectors to the bonds
exact, the spiralling would be absent on condition that expression
(\ref{eq:calD}) contained vectors $\mathbf{b}$ with real
coordinates. However, the expression depends on the exchange ones,
and, consequently, ${\cal D}$ is a linear combination of
differences $x_\mathrm{exch}-x_\mathrm{real}$ with coefficients from the DM
vector components. Therefore, the exact numerical values of
$x_\mathrm{exch}$ can strongly affect the spiralling, determining both
pitch and sense of the magnetic helix.

In table~\ref{bexch}, the calculated
with the use of (\ref{eq:coord}) lengths of the ``exchange'' bonds
$\mathbf{b}_\mathrm{e}$ are listed. As one can see, the lengths $|\mathbf{b}_\mathrm{e}|$ of
the 2nd and the 3rd bonds are considerably less than the real ones. It
rather justifies the approximation of strong tetrahedra
\cite{JansonDMI2013}, in which $\mathbf{b}_2=\mathbf{b}_3=0$.
Another important factor is the angle between the DM vector and the
bond $\mathbf{b}_\mathrm{e}$ in exchange coordinates. 
From table~\ref{bexch} it is seen that, in accordance with
\cite{GongXiang2012}, the angles $\angle(\mathbf{b}_\mathrm{r},\mathbf{D})$
are really close to $90^\circ$. But, in exchange coordinates, the
differences of $\angle(\mathbf{b}_\mathrm{e},\mathbf{D})$ from the right angle is
more considerable, particularly for the 2nd and the 3rd bonds.
 
\begin{table}[h]
\caption{\label{bexch} The changing of bond lengths and directions,
when replacing the real coordinates by the exchange ones. The energetic
parameters of bonds $\mathbf{b}_1,\ldots,\mathbf{b}_5$ are listed
in \cite{GongXiang2012} and table~\ref{5bonds}. The real
($\mathbf{b}_\mathrm{r}$) and ``exchange'' ($\mathbf{b}_\mathrm{e}$) bonds are calculated by the substitution of the coordinates from \cite{Effenberger1986} and equation (\ref{eq:coord}), respectively.}
\begin{indented}
\item[]\begin{tabular}{@{}lccccc}
\br
 & $|\mathbf{b}_\mathrm{r}|$ & $\angle(\mathbf{b}_\mathrm{r},\mathbf{D})$ & $|\mathbf{b}_\mathrm{e}|$ & $\angle(\mathbf{b}_\mathrm{e},\mathbf{D})$ & $\angle(\mathbf{b}_\mathrm{e},\mathbf{b}_\mathrm{r})$ \\
\mr
$\mathbf{b}_1$ & 0.3422 & 94.3$^\circ$ & 0.6340 & 99.5$^\circ$ & 7.8$^\circ$ \\
$\mathbf{b}_2$ & 0.3417 & 85.3$^\circ$ & 0.1053 & 71.3$^\circ$ & 25.5$^\circ$ \\
$\mathbf{b}_3$ & 0.3615 & 99.5$^\circ$ & 0.1599 & 110.2$^\circ$ & 13.2$^\circ$ \\
$\mathbf{b}_4$ & 0.3702 & 86.3$^\circ$ & 0.5443 & 83.8$^\circ$ & 11.0$^\circ$ \\
$\mathbf{b}_5$ & 0.7114 & 85.8$^\circ$ & 0.6325 & 86.9$^\circ$ & 5.0$^\circ$ \\
\br
\end{tabular}
\end{indented}
\end{table}

The significant difference between experimental and calculated values of propagation number can result from (i) an inaccuracy of {\it ab initio} calculations of the bond parameters $J_{ij}$ and $\mathbf{D}_{ij}$, (ii) the neglect of other bonds from the second and next magnetic shells, (iii) the presence of a non-isotropic interaction different from Dzyaloshinskii--Moriya one. It is important here to attend to the calculation accuracy of DM vector projections on the bond directions (in exchange coordinates), because only these projections affect the twist. If vectors $\mathbf{D}_{ij}$ are almost perpendicular to bonds, then their projections on the bond directions are relatively small in size, and even a slight change of their values can have a strong influence on the phenomenological constant ${\cal D}$ and consequently on the propagation number $k$. On the other hand, the isotropic parameters $J_{ij}$ can provide their influence both through the phenomenological constant ${\cal J}$ and the exchange coordinates. It is noteworthy, that although $J_{ij}$ and $\mathbf{D}_{ij}$ decrease with distance, their contributions into the macroscopic parameters ${\cal J}$ and ${\cal D}$, accordingly to (\ref{eq:calJ}) and (\ref{eq:calD}), are proportional to $b_{ij}^2$ and $b_{ij}$, correspondingly. Moreover, the number of neighbours also increases with distance. Thus, for example, there are eight non-equivalent bonds in the second magnetic environment ranging from 5.35 to 5.54 \AA.

\begin{figure}[h]
\begin{center}
\includegraphics[width=7cm]{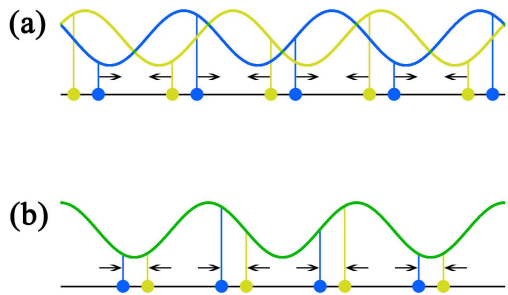}
\caption{\label{fig:exch} (a) In the real spin structure
there are several magnetic helices, corresponding to different magnetic
atoms in the unit cell, with phase shifts between them. (b) When the atoms
``move'' (arrows) from the real positions to the fictitious
(exchange) ones, the phase shifts disappear and the helices become
confluent.}
\end{center}
\end{figure}

As can be seen from (\ref{eq:coord}), the maximal deviation of
the exchange coordinate from the real one is for $x_\mathrm{II}$,
$|x_{\mathrm{II},\mathrm{exch}}-x_{\mathrm{II},\mathrm{real}}|=0.14$. In \cite{Chizhikov2013} it is
shown that for itinerant magnetics of the MnSi-type in the frame
of RKKY model, $x_\mathrm{exch}$ can have practically any value (we would
remind that the ideal coordinates are nothing but some functions
of the exchange parameters $J_{ij}$). Nevertheless, some
limitation can result from the condition of validity of the
theory. In order to study this we should first understand the
origin of the exchange coordinates. When a transition is
performed from discrete spins to continuous magnetic moment, some
frustrations appear from the fact that the spins correspond poorly
to the values of smooth function (e.g. helicoid) in their physical
positions. It is just the cause why, accordingly to equation
(\ref{eq:u1}), two kinds of canting can be distinguished: the
first one determined by the DM interaction, and the second one
connected with spatial derivatives. The latter is due to existence
of several magnetic helices, with small phase shifts from
the average single helix. The appropriate displacement of the
magnetic atoms into the fictitious positions (which does not influence the magnetic energy) makes all the individual helices confluent, and the
canting due to spatial derivatives disappears
(figure~\ref{fig:exch}). In fact, the small deviation $x_\mathrm{exch}-x_\mathrm{real}$ of the
exchange coordinate from the real one for Cu$_2$OSeO$_3$ means that the phase shifts of the individual helices are also small, which correlates with the assumption about small cantings between
neighbouring spins. Greater phase shifts would mean that we could
not already consider the spins as weak non-collinear, and the
simplifications of section~\ref{sec:collinear} would be impossible.

Another important property is the sense of the magnetic chirality.
In the case when a twist arises from a spontaneous break of
symmetry, the helix sense can be plus or minus with equal
probability and in each case it is casual. In contrast, for the
magnetics without centre of inversion, exemplified by MnSi,
Cu$_2$OSeO$_3$ and other crystals, the sense of the magnetic
chirality correlates with the structural one. Thus, in pure MnSi,
the left-handed atomic structure results in the
left-handed magnetic helix \cite{Tanaka1985,Ishida1985}. Besides,
the chiral interlink in MnSi-type helimagnetics is found to be
strongly dependent on their atomic composition
\cite{Chizhikov2013,Dmitriev2012,Grigoriev2013,Shibata2013}. In
the present work, using the magnetic data from
\cite{GongXiang2012}, we obtain $k = 0.189$, i.e. the helicoid
is right for the structure described in \cite{Effenberger1986}
and used in simulations of \cite{GongXiang2012}. Recently, this important result has been experimentally proved by Dyadkin {\it et al.} \cite{Dyadkin2014}.

In the first approximation, the cantings, being only small
corrections to spins, are perpendicular to the latters. Consequently, the cantings of
all $N$ magnetic atoms in the unit cell are determined by $2N$
variables. Thus, for Cu$_2$OSeO$_3$ the number of canting
components amounts to 32 per unit cell. Surprisingly, that, owing
the (tetrahedral) symmetry of the crystal, we can make with a
considerably lower number of parameters in order to describe the
local magnetic structure: four exchange coordinates $x_{\mathrm{I},\mathrm{exch}}$,
$x_{\mathrm{II},\mathrm{exch}}$, $y_{\mathrm{II},\mathrm{exch}}$, $z_{\mathrm{II},\mathrm{exch}}$, and four tilt vector
components $\rho_{\mathrm{I},x}$, $\rho_{\mathrm{II},x}$, $\rho_{\mathrm{II},y}$,
$\rho_{\mathrm{II},z}$. For MnSi-type helimagnetics there are only two parameters: $x_\mathrm{exch}$ and $\rho_x$.

In (\ref{eq:rho2}), the
values of tilt vectors $\boldsymbol{\rho}$ are calculated with
the use of energetic parameters $J$, $\mathbf{D}$ from
\cite{GongXiang2012}. In addition to the propagation number
$k$, given by (\ref{eq:k}), these, characterizing the canting,
vectors are the experimentally observed parameters of the
magnetic structure. In \cite{Dmitrienko2012} the conditions
were proposed of a diffraction experiment to find the canting in
the MnSi-type helimagnets. Because the magnetic atoms in these
crystals are situated at the special positions $4a$, the tilt
vectors are defined by only one parameter $\rho_x$, which is
proposed to be found in the measurement of the ``forbidden'' Bragg
reflection $00\ell$, $\ell=2n+1$ in the unwound by magnetic field
structure. In the general case of cubic helimagnets with almost
collinear spins, the antiferromagnetic (canting) part of the
structure factor is determined by the vector
\begin{equation}
\label{eq:sfactor1} \boldsymbol{\Phi} = \sum_i c_i
\boldsymbol{\rho}_i \exp( \rmi \mathbf{Q} \cdot \mathbf{r}_i ) ,
\end{equation}
where $\mathbf{Q} = 2\pi (h k \ell)$ is a reflection vector,
and summation is taken over all magnetic atoms in the unit cell.
In particular, for the copper atoms in the general positions $12b$
and the pure magnetic reflection $00\ell$, $\ell=2n+1$, this sum
has the view
\begin{eqnarray}
\label{eq:sfactor2}
\fl \boldsymbol{\Phi} = 4 (\rho_y \cos 2 \pi \ell
x + \rho_z \cos 2 \pi \ell y + \rho_x \cos 2 \pi \ell z, \nonumber \\
\rmi \rho_z \sin 2 \pi \ell x + \rmi \rho_x \sin 2 \pi \ell y + \rmi \rho_y \sin 2 \pi \ell z, 0) .
\end{eqnarray}
It is obvious that, in order to find all components of tilt
vectors $\boldsymbol{\rho}$ for magnetic atoms in all
non-equivalent positions, several forbidden reflections with
different $\ell$ should be measured.

Notice that, as far as the
measurable parameters $k$, $\boldsymbol{\rho}$ are determined by
greater number of constants $J_{ij}$, $\mathbf{D}_{ij}$, the
problem of finding of the latter from experimental data is
unsolvable in general. At the best, only some combinations of $J$
and $\mathbf{D}$ can be calculated by inversion of equations
(\ref{eq:k}) and (\ref{eq:system-rho2}). Nevertheless, the
comparison of theoretical predictions with measurements can
confirm (or refute) reliability of \textit{ab initio}
calculations of the energetic parameters.

\ack
We are grateful to O. Janson, A.~A. Tsirlin and S.~S. Doudoukine for useful discussions and encouragement and to J.~H. Yang, H.~J. Xiang and X.~G. Gong for providing us
with details of their DFT calculations of the exchange parameters
and the DM vectors in Cu$_2$OSeO$_3$. The reported study was partially supported by RFBR, research project No. 14-02-00268 a, and by two projects of the Presidium of the Russian Academy of
Sciences: ``Matter at high energy densities; Substance under high
static compression'' and ``Diffraction of synchrotron radiation in
multiferroics and chiral magnetics''.

\end{document}